\documentclass[12pt]{article} 
\usepackage[sectionbib]{natbib}
\usepackage{array,epsfig,fancyheadings,rotating}
\usepackage[]{hyperref}  
\usepackage{sectsty, secdot}
\sectionfont{\fontsize{12}{14pt plus.8pt minus .6pt}\selectfont}
\renewcommand{\theequation}{\thesection\arabic{equation}}
\subsectionfont{\fontsize{12}{14pt plus.8pt minus .6pt}\selectfont}

\usepackage{amsmath}
\usepackage{amssymb}
\usepackage{amsfonts}
\usepackage{multirow}
\usepackage{amsthm}
\usepackage{xcolor}
\usepackage{booktabs}
\usepackage{setspace}
\usepackage{xr}
\usepackage{algorithm}
\usepackage{algorithmic}
\newtheorem{theorem}{Theorem}
\newtheorem{lemma}{Lemma}
\newtheorem{corollary}{Corollary}

\newtheorem{assumption}{Assumption}
\theoremstyle{definition}
\newtheorem{definition}{Definition}

\newtheorem{remark}{Remark}
\newcommand{\myref}[1]{(\ref{#1})}

\begin{document}
	
	
	\renewcommand{\baselinestretch}{1}
	
	\markright{ \hbox{\footnotesize\rm Statistica Sinica
		}\hfill\\[-13pt]
		\hbox{\footnotesize\rm
		}\hfill }
	
	\markboth{\hfill{\footnotesize\rm Soumik Purkayastha AND Peter X.-K. Song} \hfill}
    {\hfill {\footnotesize\rm Quantification and Inference of Asymmetric Relations under GEMs} \hfill}
	
	\renewcommand{\thefootnote}{}
	$\ $\par
	
	
	\fontsize{12}{14pt plus.8pt minus .6pt}\selectfont \vspace{0.8pc}
\centerline{\large\bf Quantification and Inference of Asymmetric Relations  }
\vspace{2pt} 
\centerline{\large\bf Under Generative Exposure Mappings}
\vspace{.4cm} 
\centerline{Soumik Purkayastha \\ Peter X.-K. Song} 
\vspace{.4cm} 
\centerline{\it University of Pittsburgh \\ University of Michigan}
 \vspace{.55cm} \fontsize{9}{11.5pt plus.8pt minus.6pt}\selectfont


\begin{quotation}
\noindent {\it Abstract:} Learning directionality between variables is crucial yet challenging, especially for mechanistic relationships without a priori ordering assumptions. We propose a coefficient of asymmetry to quantify directional asymmetry using Shannon's entropy within a generative exposure mapping (GEM) framework. GEMs arise from experiments where a generative function $g$ maps exposure $X$ to outcome $Y$ through $Y = g(X)$, extended to noise-perturbed GEMs as $Y = g(X) + \epsilon$. Our approach considers a rich class of generative functions while providing statistical inference for uncertainty quantification—a gap in existing bivariate causal discovery techniques. We establish large-sample theoretical guarantees through data-splitting and cross-fitting techniques, implementing fast Fourier transformation-based density estimation to avoid parameter tuning. The methodology accommodates contamination in outcome measurements. Extensive simulations demonstrate superior performance compared to competing causal discovery methods. Applied to epigenetic data examining DNA methylation and blood pressure relationships, our method unveils novel pathways for cardiovascular disease genes \emph{FGF5} and \emph{HSD11B2}. This framework serves as a discovery tool for improving scientific research rigor, with GEM-induced asymmetry representing a low-dimensional imprint of underlying causality.

\vspace{9pt}
\par
\end{quotation}\par

	\def\thefigure{\arabic{figure}}
	\def\thetable{\arabic{table}}
	
	\renewcommand{\theequation}{\thesection.\arabic{equation}}

	\fontsize{12}{14pt plus.8pt minus .6pt}\selectfont
	
	\section{Introduction}  
\label{sec:intro}
 In many statistical applications, \textit{ordering} among variables is prefixed according to a certain scientific hypothesis, scientific knowledge, or a problem of interest. However, when the ordering itself is of scientific interest or if statistical analyses are sensitive to the choice of ordering, an inevitable challenge lies in inferring a sense of order in a given set of variables.  We posit that the notion of order, or \textit{asymmetry} (if it exists), is reflective of an underlying generative mechanism that maps an exposure $X$ to an outcome $Y$. This paper aims to develop statistical methods that quantify and infer said asymmetry in a general setup
for continuous data with or without contamination of outcomes. We term this general setup as the \emph{generative exposure mapping} (GEM) {that defines an underlying ordering}.  Conceptually, GEMs arise from experiments where a mechanistic procedure, or generative function (GF) $g$ outputs an outcome $Y$ for an input exposure $X$ through the mapping $Y = g(X)$. To accomodate contaminated, or noise-perturbed outcomes, the GEM framework can be expanded to \emph{noise-perturbed GEM} (NPGEM), denoted by $Y = g(X) + \epsilon$, where $\epsilon$ is treated as the outcome contaminant. 
In this work, we show that an effective strategy for the assessment of asymmetry in GEMs (with mild structural assumptions on GFs) is to prove or disprove the ordering induced from a hypothesized GEM using Shannon's information theory. We propose a conceptually easy and computationally manageable statistical methodology to quantify and provide inference for relational asymmetry. As evidenced in this paper, the utility of Shannon's information theoretic measures allows us to study asymmetry with no need of estimating the GF $g$.

The simplest occasions of asymmetry involve placing spatial or temporal ordering conditions between the two \citep{cox1992causality}.  Establishing a presumed causal ordering of variables often requires using specific subject-matter knowledge in connection to external or \textit{a priori} information \citep{cox1990role}. Alternatively, distributional asymmetries are studied by factorizing the joint density $f_{{XY}}$ as the product of the marginal $f_{{X}}$ and the conditional $f_{{Y}|{X}}$; see \cite{choi2020, tagasovska2020, ni2022bivariate}, among others. 
Recently, some asymmetric measures of association have been proposed, including the generalized measure of correlation~\citep{zheng2012generalized} and a rank-based asymmetry measure~\citep{Chatterjee2020}, but none have meaningful connections to directionality implied by GEMs. For example, in \cite{zheng2012generalized} the authors propose generalized measures of correlation for asymmetry and nonlinearity but they cannot track directionality implied by GEMs. 
Similarly, \cite{Chatterjee2020} proposes a rank-based measure of association that is asymmetric, but fails to reflect any generative mechanism. 
The role of ordering in causal inference is apparent where the existence of \textit{a priori} direction of causation is inherently hypothesized. This prompts the need for \textit{causal discovery}  on learning underlying causal structures from observational data \citep{pearl2009}. 

We establish a self-contained theoretical framework that significantly broadens the existing Information Geometric Causal Inference (IGCI) \citep{daniusis2012inferring, janzing2012information} framework for examining asymmetry in exposure-outcome pairs. In the IGCI framework, the authors propose methodology to examine directionality in the GEM $Y = g(X)$ for bijective $g$. The GEM framework broadens the scope of this framework by considering a much richer class of GFs, including bijective functions. Moreover, we provide statistical inference to quantify the uncertainty in the determination of underlying directionality, while most existing bivariate causal discovery techniques \citep{hoyer2008, Fonollosa2019,  Blbaum2019} do not. Moreover, our method does not assume any sort of dependence structure between the exposure $X$ and contaminant $\epsilon$ in NPGEMs, providing a generalization of another causal discovery framework proposed by \cite{hoyer2008}.

We organize the paper as follows: Section \ref{sec:setup}  introduces the setup of GEMs and some basic information theoretic concepts. Section \ref{sec:strong} provides a recap of IGCI and provides a population-level measure of asymmetry under a hypothesized GEM. Section \ref{sec:strong_cont} introduces NPGEMs to allow for errors in the outcome $Y$ with technical justifications for both feasibility and robustness of the proposed coefficient of asymmetry with random samples from the GEM. 
Section \ref{sec:estim_inf} presents estimation and inference details; we implement a fast Fourier transformation-based density estimation technique to estimate key estimands of interest under a proposed GEM, followed by a cross-fitting technique to quantify estimation uncertainty while improving statistical efficiency. Finally,  Sections \ref{sec:sims} and \ref{sec:rda} exhibit the performance of the proposed framework, estimation, and inference through simulation studies and real data applications respectively. 

\section{Preliminaries}
\label{sec:setup}
We begin by outlining basic concepts and notation used throughout our paper. We first define the settings of our GEM and then introduce some elementary information theoretic concepts that are key to our framework. 
\subsection{The generative exposure mapping setup}
\label{subsec:gcm_setup}
Let us consider a population $\mathcal{U}$ of units indexed by $i = 1, \ldots n$ on which a randomized experiment is performed. Since an experimental design details the selection of a particular exposure value $x$ from the support $\mathcal{X}$ of continuous exposure $X$, we can say the probability density function (PDF) of $X$, given by $f_X$, governs the experiment. 
Let the $i-$th population unit independently get exposure $X_i$. According to \cite{cox1990role, cox1992causality}, a causal ordering between $X$ and outcome $Y$ must be explained through (i) a mechanism governing the exposure and (ii) a generative process yielding the outcome, with exposure as the input. We consider a structured generative mechanism specified by the GEM of the form with an ordering between two variables: 
\begin{equation}
\label{eq:ANM}
    {Y} = g({X}).
\end{equation} Let $Y_i = g(X_i)$ be the outcome of the  $i$-th population unit in $\mathcal{U}$, which are assumed to be independent. In a GEM, note that the behaviour of $Y$, described by its PDF $f_Y$, is dictated entirely by $f_X$ and $g$. The exposure mapping model of \cite{aronow2017} is noted to be a special case of the GEM in  \myref{eq:ANM} for discrete $X$ and is further equivalent to the ``effective treatments'' setup described by \cite{Manski2013}. The overall challenges of examining asymmetry in GEMs involve inferring whether exposure $X$ yields outcome $Y$ or \textit{vice-versa} from paired observations $\left\{ ({x}_i, {y}_i) \right\}_{i=1}^n$.  We focus exclusively on assessing asymmetry in GEMs by leveraging information-theoretic notions of \textit{entropy}. 
The subsequent section provides a brief review of entropy, which will play a key role in quantifying asymmetry in GEMs. It is important to note that our framework presupposes that $X$ and $Y$ are associated, which must be established before testing for directionality. If needed, one can use the methodology proposed by \cite{purkayastha2023} to test for association using mutual information \citep{Cover_2005}.

\subsection{Basic information theoretic concepts}
\label{subsec:infotheo}
Let $X$ and $Y$ be two random variables with joint density function $f_{XY}$. Let $f_X$ and $f_Y$ be the marginal densities of $X$ and $Y$, respectively. Further, let $\mathcal{X}$ and $\mathcal{Y}$ denote the respective support sets of $X$ and $Y$. 
The \textit{joint differential entropy} of $(X, Y)$ and \textit{marginal differential entropy} of $X$ are given by $H(X, Y) = E_{XY}\left\{ - \log f_{XY}(X, Y) \right\}$ and $H(X) = E_{X}\left\{ - \log f_{X}(X) \right\}$ respectively \citep{Shannon1948}. Differential entropy measures the randomness of a continuous random variable \citep{Orlitsky2003} and is a limiting case of Shannon's entropy, which was originally described for discrete random variables. For the remainder of this paper, we omit the word differential, although our focus is always continuous random variables. 
In the next section, we demonstrate the flexibility and capacity of Shannon's entropy measure to quantify asymmetry under GEMs. 

\section{Asymmetry in GEMs}
\label{sec:strong}
{In this section we derive a legitimate population-level measure that will be used to confirm ordering or asymmetry under a hypothesized GEM.

Let us begin with the case of a GEM as described by \myref{eq:ANM}; intuitively, we consider the exposure $X$ is collected from an experiment that is governed by a density $f_X$. This setup is often used in Fisher's fiducial inference~\citep{Hannig} in which errors, although unobservable in practice, may be simulated from a certain pivotal distribution. With $X$ serving as an input, the \textit{GF} $g$ generates the population-level outcome $Y = g(X)$, which has density $f_Y$. We will consider a specific classification of \textit{GF}s, denoted by $\mathcal{G}_{-}$, as defined below. 
\begin{definition} \label{def:contract}
  Let $\mathcal{G}_{-}$ be a class of differentiable functions on a compact set $\gamma \subset \mathbb{R}$. For any $g \in \mathcal{G}_{-}$, let the following integral be well-defined and satisfy the inequality
  \begin{equation*}
      \exp \left({|\gamma|}^{-1} \int_\gamma \log |\nabla g(x)|\, d x\right)<1, 
  \end{equation*}
  where $|\gamma|$ denotes the Lebesgue measure of $\gamma$, and $\nabla$ denotes the gradient of $g$. We interpret any function $g \in \mathcal{G}_{-}$ to exhibit \emph{contracting dynamics} over the support $\gamma$, meaning that the {geometric mean} of $|\nabla g(x)|$ over $\gamma$ is strictly less than one.
\end{definition}

\begin{remark} \label{rem:dynamics}
    Alternatively, we may have a class of functions $\mathcal{G}_{+}$ with `expanding dynamics' over support $\gamma$; for any $g^\star \in \mathcal{G}_{+}$, we have
    \begin{equation*}
      \exp \left( {|\gamma|}^{-1} \int_\gamma \log |\nabla g^\star(x)| d x\right)>1, 
  \end{equation*} i.e., the {geometric mean} of $|\nabla g(x)|$ over $\gamma$ is strictly more than one. 
\end{remark}

An identifiability assumption required to unearth the induced asymmetry from a GEM is that the distribution of exposure $X$ (or the law of $X$) and the mechanism of the \textit{GF} $g$  do not influence each other when yielding the outcome $Y$, which we quantify through functional orthogonality, as defined below. In the literature of functional analysis, orthogonality is appropriate to characterize the notion of ``no influence'', which will be adopted in this paper. Similar assumptions are found in \cite{Janzing2008, daniusis2012inferring, mooij_2016}. 
\begin{assumption}
    \label{ass:Iden1} 
Let $g$ be a differentiable function and $f_X$ be the density that governs an experiment with exposure $X$ with compact support $\mathcal{X}$. We say that $g$ and $f_X$ are functionally orthogonal if they satisfy the equality $$\int_{\mathcal{X}}  \log \left(\lvert \nabla g(x) \rvert \right)f_X(x)dx = \lvert\mathcal{X}\rvert^{-1} \int_{\mathcal{X}}\log \left(\lvert \nabla g(x) \rvert \right)dx,$$
where the $\nabla$ operator denotes the gradient of $g$ with respect to its argument.
\end{assumption}
\begin{remark}
\label{rem:strong_1}
Assumption \ref{ass:Iden1} automatically holds when $X \sim \mathcal{U}(0,1)$. This is analogous to the assumption of randomization or no confounding in the school of Neyman-Rubin causality; the uniform distribution on $X$ is analogous to randomization in an experiment that leads to no bias in operating exposure $X$.
\end{remark}

\begin{remark}
Assumption~\ref{ass:Iden1} may also hold under non-uniform distributions such as Beta$(a,b)$, provided there is a functional balance between the shape of $f_X$ and the transformation $g$. For example, if $g(x) = x^k, k \neq 1$ or $g(x) = \log (x)$, Assumption~\ref{ass:Iden1} holds for specific choices of $(a,b)$. Intuitively, this resembles balancing moments or information content: the shape of $f_X$ counteracts the local expansion or compression induced by $g$, yielding an overall uncorrelatedness. As a concrete case, when $g(x) = x^k$, Assumption~\ref{ass:Iden1} holds for Beta$(a,b)$ distributions with $\mathbb{E}[\log X] = -1$, which includes asymmetric shapes such as Beta$(1,1.5)$. This highlights that functional orthogonality is not limited to uniform inputs, but rather depends on a compatibility between the distribution of $X$ and the dynamics of $g$.
\end{remark}

\begin{remark}
To interpret Assumption \ref{ass:Iden1}, consider the quantities $\log \left(\lvert \nabla g(X) \rvert \right)$ and $f_X(X)$ as random variables on support $\mathcal{X}$. Then, their covariance with respect to the uniform distribution on $\mathcal{X}$ is proportional to 
  \begin{equation*}
    \int_{\mathcal{X}}  \log \left(\lvert \nabla g(x) \rvert \right)f_X(x)dx - \lvert\mathcal{X}\rvert ^{-1} \int_{\mathcal{X}} \log \left(\lvert \nabla g(x) \rvert \right)dx \int_{\mathcal{X}}  f_X(x) dx, 
\end{equation*} where the last integral is equal to one. Consequently, the equality in \ref{ass:Iden1} expresses that the density of exposure, given by $f_X(x)$, and the (log-transformed) dynamics of the generative mechanism, given by $\nabla g(x)$, are uncorrelated mechanisms of nature, linking with the notion of algorithmic independence \citep{Janzing2008}. {Geometrically, Assumption \ref{ass:Iden1} requires that the parts of the input domain where the data are densely distributed are not systematically related (or, correlated) to the parts of the function $g$ that are most expansive or contractive (i.e., where $\lvert \nabla g(x)\rvert$ is large or small). This holds naturally in a randomized experiment where $X$ is, for example, uniformly distributed.}
\end{remark}

{The validity of Assumption \ref{ass:Iden1} in the absence of unmeasured confounding may be checked; we provide an a practical guide in Algorithm \ref{alg:ortho_check} below. In brief, we outline steps to numerically estimate the left and right side quantities in Assumption \ref{ass:Iden1} and obtain the difference between them. We call this estimate the ``orthogonality deviation score'' and denote it by $D$. In addition to the the point estimate $\hat{D}$, we also provide a bootstrap confidence interval. A confidence interval that is narrow and contains to zero suggests Assumption 1 is plausible for the specified direction. A confidence interval that is far from zero indicates a likely violation of the assumption. Please see section 5 of supplement for an example illustrating two contrasting scenarios, one satisfying the assumption and one violating it.}
\newenvironment{algocolor}{%
  \color{black} 
}{}
\begin{algorithm}
\begin{algocolor}
\caption{{Orthogonality Deviation Score ($D$) for Assumption \ref{ass:Iden1}}}
\label{alg:ortho_check}
\begin{algorithmic}[1]
\STATE \textbf{input:} dataset of paired observations $\{(x_i, y_i)\}_{i=1}^n$. 
\STATE \textbf{output:} $D$ and its 95\% bootstrap confidence interval. 
\STATE fit a smoothing spline to obtain an estimate of $\hat{g}(x)$, numerically compute $\nabla\hat{g}(x)$.
\STATE estimate $\text{LHS}_{\text{est}} \gets n^{-1} \sum_{i=1}^{n} \log(|\nabla \hat{g}(x_i)|)$
\STATE estimate $\text{RHS}_{\text{est}} \gets |\hat{\mathcal{X}}|^{-1} \int_{\min(X)}^{\max(X)} \log(|\nabla \hat{g}(x)|) dx$
\STATE calculate $D \gets \text{LHS}_{\text{est}} - \text{RHS}_{\text{est}}$.
\FOR{$b = 1$ to $B$ (e.g., $B=500$)}
    \STATE generate a bootstrap sample $\{(x_i^*, y_i^*)\}_{i=1}^n$.
    \STATE repeat steps 3-7 on the bootstrap sample to compute a score $D_b^*$.
\ENDFOR
\STATE construct a 95\% CI from the empirical distribution of scores
\RETURN $\hat{D}$ and its 95\% bootstrap Ci.
\end{algorithmic}
\end{algocolor}
\end{algorithm}

We now propose the decision rule that will detect the direction of causation $X \rightarrow Y$ induced by the GEM $Y = g(X)$. The decision rule uses the following asymmetry coefficient between $X$ and $Y$, as defined below.
\begin{definition}
\label{def:pac_alt}
    Let $g \in \mathcal{G}_{-}$ be a \textit{GF} in \myref{eq:ANM}. Then, we define the \textit{asymmetry coefficient} between $X$ and $Y$ as follows $$C_{X \rightarrow Y} := H(X) - H(Y),$$ where $H(X)$ and $H(Y)$ are the entropy functions of $X$ and $Y$ respectively. 
    \end{definition} \noindent The key challenge of this approach lies in the fact that we do not have knowledge of the generative function $g$. However, we see that the {asymmetry coefficient} is able to capture generative asymmetry even without explicitly knowing the form of $g$, since we have the following
\begin{equation*} \label{eq:igci}
    \begin{aligned}
        C_{X \rightarrow Y} &=  -\int_{\mathcal{X}}  \log \left(\lvert \nabla g(x) \rvert \right)f_X(x)dx = -\lvert\mathcal{X} \rvert ^{-1} \int_{\mathcal{X}}  \log \left(\lvert \nabla g(x) \rvert \right)dx > 0, 
    \end{aligned}
\end{equation*} where the inequalities are supported by Assumption \ref{ass:Iden1} and Definition \ref{def:contract}. Following Definition \ref{def:pac_alt}, we formulate the following hypothesis to confirm or negate the presence of the pathway $X \rightarrow Y$ for a given GEM $Y = g(X)$, where $g \in \mathcal{G}_{-}$ has contracting dynamics: 
\begin{equation} \label{eq:pointwise}
\begin{aligned}
    H_{0}:  C_{X \rightarrow Y} \leq 0 \quad &\text{ vs }\quad  H_{1}: C_{X \rightarrow Y} > 0. 
\end{aligned}
\end{equation} If $H_{0}$ is rejected, we obtain evidence to support presence of the pathway $X \rightarrow Y$ induced by the GEM in \myref{eq:ANM}. 
\begin{remark} \label{rem:expand}
    The positive sign of $C_{X \rightarrow Y}$ relies on the `contracting dynamics' nature of $g$. However, in the scenario where $g$ has expanding dynamics (see Remark \ref{rem:dynamics}), we will have $C_{X \rightarrow Y} < 0$. Hence, we must modify hypothesis in (\ref{eq:pointwise}) accordingly, given by 
    \begin{equation} 
    \label{eq:pointwise2}
    H_{0}: C_{X \rightarrow Y} \geq 0 \quad \text{ vs }\quad  H_{1}: C_{X \rightarrow Y} < 0. 
    \end{equation}
\end{remark}

\begin{remark}
{A key challenge from a practical standpoint is that if $g$ has expanding dynamics, its inverse $g^{-1}$ must have contracting dynamics; please see section 5 of the supplement for a formal proof. While one {asymmetry coefficient} is defined as $C_{X\rightarrow Y} = H(X) - H(Y)$, while the coefficient for the reverse direction is $C_{Y\rightarrow X} = H(Y) - H(X) = -C_{X\rightarrow Y}$. Consider the two hypotheses:
\begin{enumerate}
    \item {Setting A:} The true causal direction is $X \rightarrow Y$ with an {expanding} function $g$. This implies $C_{X\rightarrow Y} < 0$.
    \item {Setting B:} The true causal direction is $Y \rightarrow X$ with a {contracting} function $h$. This implies $C_{Y\rightarrow X} > 0$.
\end{enumerate}
The mathematical conditions for these two settings, $C_{X\rightarrow Y} = H(X) - H(Y) < 0$ and $C_{Y\rightarrow X} = H(Y) - H(X)> 0$, are identical. Hence, just by observing the sign of the asymmetry coefficient, we cannot distinguish between two complementary scenarios. The key to resolving this ambiguity is that Assumption 1 is not symmetric: it can be proved that at most one of the two settings described above will be compatible with Assumption 1. In other words, Assumption 1 will be compatible with at most one of the implied causal directions $X \rightarrow Y$ or $Y \rightarrow X$. Please see section 5 of the supplement for a formal proof.}
\end{remark}

\begin{remark}
\label{rem:gen_fun_verify}
{If we have {prior domain knowledge} that the generative mechanism is, for instance, known to be {expanding}, the ambiguity is resolved immediately. Upon observing a significantly negative asymmetry coefficient ($C_{X \rightarrow Y} < 0$), we can confidently conclude the direction is $X \rightarrow Y$. If we observe a positive coefficient, we would conclude the direction is $Y \rightarrow X$. If we do not have prior domain knowledge, we propose a exploratory tool to detect if the generative function in the GEM $Y = g(X)$ is expanding or contracting; please see section 6 of the supplement for more details. }
\end{remark}

{Further, it is worth pointing out that even when Assumption 1 does not hold, the proposed analytics and algorithms in our framework remain useful and produce meaningful, yet weaker interpretability of the directionality. That is, although without Assumption 1 the interpretability of causality cannot be guaranteed, the detection of relational directionality remains meaningful and useful in the study of asymmetric relationships. This can easily be seen by noting that $H(X) - H(Y) = H(X|Y) - H(Y|X)$. Therefore, $H(X) > H(Y)$  implies $H(X|Y) > H(Y|X)$. This indicates that the uncertainty of $Y$ given $X$ is less than the uncertainty of $X$ given $Y$, establishing that $X$ provides more information about $Y$ than vice-versa.}

\section{Asymmetry in the presence of noise perturbation}
\label{sec:strong_cont}
In a practical data generation scenario, measurement errors or noise perturbations are inevitable. Thus, we now extend the setup of the previously noise-free GEM by adding randomness on the outcome $Y$, so a noise-perturbed GEM (NPGEM) takes the form:
\begin{equation} \label{eq:ANM_eps}
    Y^* = g(X) + \epsilon, \text{ such that } \mathbb{E}(\epsilon) = 0, \mathbb{V}(\epsilon) = \sigma, 
\end{equation} where $Y^*$ is the `contaminated' version of outcome $Y$. 

\subsection{Contamination for GFs with contracting dynamics}
From Definition \ref{def:pac_alt} and the hypothesis in \ref{eq:pointwise}, note that the asymmetry coefficient can measure population-level asymmetry between exposure and outcome for GFs with contracting dynamics even with noise perturbation when we encounter $Y^*$ instead of $Y$ in the population-level model, so long as the ordering of $H(Y^*)$ being lower than $H(X)$ is preserved. 

We introduce an intermediate  NPGEM with normally distributed disturbance $Z$ having mean $0$ and variance $1$, such that $H(Y^*) \leq H(Y + \sqrt{\sigma^\prime}Z)$ for a certain $\sigma^\prime >0$. This NPGEM is merely for a technical need; de Bruijn's theorem \citep{Cover_2005} has shown that $H(Y + \sqrt{\sigma^\prime}Z)$ is an increasing function of $\sigma^\prime$. We state the following theorem: 
\begin{theorem}
\label{lem:robustness}
We consider a NPGEM given by \myref{eq:ANM_eps} with $g \in \mathcal{G}_{-}$,  $f_Y$ and $H(Y)$ denoting the density and entropy of $Y$ respectively. Then, the upper-bound on the entropy of $Y^*$ is given by $H \left(Y^*\right) \leq H(Y) + \frac{1}{2} \log \left(\sigma^\prime I(Y) + 1 \right),$ where $\sigma^\prime$ is a value such that $H(Y^*) \leq H(Y + \sqrt{\sigma^\prime} Z)$ with $Z \sim N(0, 1)$ and $X \perp Z$, and $I(Y) := \mathbb{E}_Y \left[ \nabla_Y \log \left(f_Y(Y) \right) \right]^2$ is the (nonparametric) Fisher information of $Y$. 
\end{theorem}
Note that the existence of the $\sigma^\prime$ is ensured by de Bruijn’s identity. We refer the reader to the supplement for the proof of Theorem \ref{lem:robustness}. To ensure the asymmetry coefficient properly measures the induced asymmetry, we must have  $H \left(Y^*\right) \leq H(Y) +  \log \left(\sigma^\prime I(Y) + 1 \right)/2 < H(X)$, implying
\begin{equation}
\label{eq:s2n}
    I(Y)< \frac{\exp(2C_{X \rightarrow Y}) - 1}{\sigma^\prime} = \frac{\exp(2C_{X \rightarrow Y}) - \exp(2C^0_{X \rightarrow Y})}{\sigma^\prime}, 
\end{equation} where $C^0_{X \rightarrow Y} \equiv 0$ is value of $C_{X \rightarrow Y}$ in a \textit{balanced} or \textit{symmetric} relation. We argue that the numerator $\exp(2C_{X \rightarrow Y}) - \exp(2C^0_{X \rightarrow Y})$ is a measure of the deviation from symmetry and serves as the signal that our method must capture. 
To arrive at \myref{eq:s2n}, note that we have (i) $C_{X \rightarrow Y} = H(X) - H(Y) > 0$, and (ii) Theorem \ref{lem:robustness} ensures $\sigma^\prime I(Y) < \exp(2C_{X \rightarrow Y}) - \exp(2C^0_{X \rightarrow Y}).$ Hence, the asymmetry coefficient works in low-noise regimes of the NPGEMs, where \myref{eq:s2n} provides a lower bound on the signal-to-noise ratio that the asymmetry coefficient can tolerate and remains effective to quantify NPGEM-induced asymmetry.


\subsection{Contamination for GFs with expanding  dynamics}

For $Y^\star =g(X)+\epsilon$ with \textit{expanding dynamics} (\( g \in \mathcal{G}_{+} \)), we have $H(g(X)) > H(X)$.  To ensure that the  entropy of the contaminated output \( Y^\star \) remains greater than that of the input \( X \), we consider the impact of the transformation \( g \) and the added noise \( \epsilon \). If \( g \) has \textit{expanding dynamics}—that is, \( H(g(X)) > H(X) \)—then we can guarantee \( H(Y^\star) > H(X) \) under the added condition of independence of noise $\epsilon$ and exposure $X$. Using the entropy power inequality \citep{Cover_2005}, we can claim that the differential entropy of the contaminated output satisfies $H(Y^\star) \geq \max \big\{ H(g(X)), H(\epsilon) \big\} > H(X)$. Hence, adding independent noise cannot reduce the entropy below that of \( g(X) \), which already exceeds \( H(X) \) when $g \in \mathcal{G}_{+}$. 

\section{Estimation and inference}
\label{sec:estim_inf}
In the GEM framework, we must estimate and perform inference using $\hat{C}_{X \rightarrow Y}$. To do so, we must first estimate the underlying marginal densities $f_X$ and $f_Y$, which may be thought of as infinite-dimensional nuisance parameters. If the same data that were used for density estimation are also used to provide inference for ${C}_{X \rightarrow Y}$, standard inferential procedures may suffer from bias. To circumvent this, we consider a data-splitting and cross-fitting approach \citep{chernozhukov2018double}. Moreover, we use a fast Fourier transformation (FFT) technique to estimate the underlying density functions and provide plug-in estimators of entropy. This technique is key to providing a stable estimator. To the best of our knowledge, our method is the first to combine FFT-based techniques and data-splitting techniques to estimate and perform inference on entropy. 
The density estimation technique in Section \ref{subsec:density} provides an accurate and fast solution without incurring the need for tuning parameters. Based on the estimated densities, we obtain consistent estimates of $\hat{C}_{X \rightarrow Y}$. Next, in Section \ref{subsec:split_fit}, we describe the data-splitting and cross-fitting technique that allows us to provide an inference rule for testing 
for GEM-induced asymmetry  using either the hypothesis in \ref{eq:pointwise} for GFs with contracting dynamics or the hypothesis in \ref{eq:pointwise2} for GFs with expanding dynamics.

\subsection{Self-consistent density estimation}
\label{subsec:density}
The self-consistent estimator (SCE) was proposed by \cite{Bernacchia2011, OBrien2016}  to minimize the mean integrated squared error (MISE) between the estimated density and the true density without incurring any manual parameter tuning. The estimation process relies on fast Fourier transforms (FFT). Utilizing this  `optimal' density estimator,  \cite{purkayastha2023} propose a plug-in estimator of $MI$, termed as the \texttt{fastMI}, which is shown to be a consistent and fast estimator. Extending the usage of the self-consistent density estimator here, we then estimate the marginal entropies $\hat{H}(X)$ and 
$\hat{H}(Y)$, thereby obtaining an estimate of $\hat{C}_{X \rightarrow Y}$. 

Let us consider a random sample denoted by $\mathcal{S} = \{X_1, X_2, \ldots, X_n\}$ from an unknown density $f$ with support $\mathcal{X}$ (without loss of generality, $\mathcal{X} = \mathbb{R}$). We assume $f$ belongs to the Hilbert space of square-integrable functions, given by $\mathcal{L}^2 = \left\{f: \int f^2({x})d{x} < \infty \right\}.$
The SCE is denoted by $\hat{f} \in \mathcal{L}^2$. First, in order to define $\hat{f}$, we require a kernel function $\mathcal{K}$, which belongs to the class of functions given by $$\mathbb{K} := \left\{\mathcal{K}: \mathcal{K}({x}) \geq 0, \mathcal{K}({x}) = \mathcal{K}(-{x}); \int \mathcal{K}({t})d{t} = 1 \right\}.$$ Specifically, $\hat{f}$  is the convolution of a kernel $\mathcal{K}$ and delta functions centered on the dataset: 
\begin{equation}
\footnotesize
\begin{aligned}
 \label{eqn:kde}
     \hat{f}({z}) \equiv {n}^{-1} \sum_{j = 1}^{n} \mathcal{K}({x} - {X}_j) =  {n}^{-1} \sum_{j = 1}^{n} \int_{\mathbb{R}} \mathcal{K}({s}) \delta({x} - {X}_j - {s})d{s}, \quad {x} \in \mathbb{R} 
\end{aligned} 
\end{equation} where $\delta(x)$ is the Dirac delta function.
The optimal $\hat{f}$ is identified by the optimal kernel $\hat{\mathcal{K}}$, where ``optimality'' is intended as minimising the MISE between the true density $f$ and the estimator $\hat{f}$:
\begin{equation}
\begin{aligned}
 \label{eqn:mise}
     \hat{\mathcal{K}} = \underset{{\mathcal{K}} \in \mathbb{K}}{\text{argmin}} \ \text{\textit{MISE}}(\hat{f}, f)  = \underset{\mathcal{K} \in \mathbb{K}}{\text{argmin}} \  \mathbb{E}\left[\int_{\mathbb{R}}\{\hat{f}(x)-f(x)\}^{2} dx\right],
\end{aligned}
\end{equation} where the $\mathbb{E}$ operator denotes taking expectation over the entire support of $f$. The SCE in  \myref{eqn:kde} may be represented equivalently by its inverse Fourier transform pair, $\hat{\phi} \in \mathcal{L}^2$, given by $\hat{\phi}(t) = \mathcal{F}^{-1} \big(\hat{f}(z)\big)  = {\kappa}(t) \mathcal{C}(t),$ where $\mathcal{F}^{-1}$ represents the multidimensional inverse Fourier transformation from space of data $z \in \mathbb{R}$ to frequency space coordinates $t \in \mathbb{R}$. ${\kappa} = \mathcal{F}^{-1} \big(\mathcal{K}\big)$ is the inverse Fourier transform of the kernel $\mathcal{K}$ and $\mathcal{C}$ is the empirical characteristic function (ECF) of the data, defined as $\mathcal{C}(t) = {n}^{-1} \sum_{j = 1}^{n} \exp{(i t {Z}_j)}$.
\cite{Bernacchia2011} derive the optimal transform kernel $\hat{\kappa}$ minimizing the $MISE$ given by  \myref{eqn:mise}, given as follows:
\begin{equation}
\begin{aligned}
 \label{eqn:kernel_fouriertransformed}
 \hat{\kappa}(t) &= \frac{{n}}{2(n-1)} \left[1 + \sqrt{1 - \frac{4({n}-1)}{ \left| n \mathcal{C}(t) \right|^2 }} \right] I_{A_{n}}(t),
\end{aligned}
\end{equation} where $A_{n}$ serves as a low-pass filter that yields a stable estimator~\citep{purkayastha2023}. We follow the nomenclature of \cite{Bernacchia2011} and denote $A_{n}$ as the set of ``acceptable frequencies''. The optimal transform kernel $\hat{\kappa}$ in  \myref{eqn:kernel_fouriertransformed} may be anti-transformed back to the real space to obtain the optimal kernel $\hat{\mathcal{K}} \in \mathbb{K}$, which yields the optimal density estimator $\hat{f}$ according to Equation \ref{eqn:kde}. 
\begin{remark}
{
   The key components are (a) {The ECF $\mathcal{C}(t)$:} The empirical characteristic function, $\mathcal{C}(t)$, is the Fourier transform of the data and represents the observed ``signal'' in the frequency domain, and (b) {the indicator function $I_{A_{n}}(t)$:} This term acts as a ``low-pass filter''. It is active (equal to 1) only for frequencies where the signal strength, represented by $|n\mathcal{C}(t)|^2$, is above a certain threshold.  In \cite{purkayastha2023} we define {the filter set as follows:} $$A_n(t) = \left\{ t \in \mathbb{R} : |C(t)|^2 \ge C_{\min}^2 = \frac{4(n-1)}{n^2} \right\}.$$
For frequencies where the signal is too weak (i.e., likely dominated by noise), the term inside the square root would be negative; the filter sets $\hat{\kappa}(t)$ to zero for these frequencies, effectively removing them and ensuring a stable estimate. {As shown in \cite{purkayastha2023}, this} self-consistent, data-driven process determines the optimal level of smoothing automatically, in contrast to traditional kernel methods that rely on user-specified bandwidths.}
\end{remark}

Theorem \ref{thm:strong_conv} presents the sufficient conditions for the estimate $\hat{f}$ to converge to the true density $f$ for ${n} \rightarrow \infty$. First, we state the technical assumptions needed for Theorem \ref{thm:strong_conv} to hold. 

\begin{assumption}
\label{ass:th_1_1}
    Let the true density $f$ be square-integrable and its corresponding Fourier transform $\phi$ be integrable. 
\end{assumption}
\begin{assumption}
\label{ass:th_1_2}
     Let us assume the following about the low-pass filter $A_{n}$: $\mathcal{V}(A_{n}) \rightarrow \infty, \ \mathcal{V}(A_{n})/\sqrt{{n}} \rightarrow 0 \text{ and } \mathcal{V}(\bar{A}_{n}) \rightarrow 0 \text{ as } n \rightarrow \infty,$ where $\bar{A}_{n}$ is the complement of $A_{n}$ and the volume of $A_n$ is given by $\mathcal{V}(A_{n})$.
\end{assumption}
\begin{assumption}
\label{ass:th_1_3}
    Let the true density $f$ be continuous on dense support $\mathcal{X}$. 
\end{assumption}

\begin{theorem}
\label{thm:strong_conv}
If Assumptions \ref{ass:th_1_1} and \ref{ass:th_1_2} hold, then the self consistent estimator $\hat{f}$, which is defined by \myref{eqn:kde} - \myref{eqn:kernel_fouriertransformed}, converges almost surely to the true density as $n \rightarrow \infty$. Further, if Assumption \ref{ass:th_1_3} holds, we have uniform almost sure convergence of $\hat{f}$ to $f$ as $n \rightarrow \infty$. 
\end{theorem} The proof of Theorem \ref{thm:strong_conv} is omitted here; see \cite{purkayastha2023} for the detail.  

\subsection{Data-splitting and cross-fitting inference}
\label{subsec:split_fit}
We have noted earlier that if the same data that were used for density estimation are also used to provide inference for ${C}_{X \rightarrow Y}$, standard inferential procedures may suffer from bias \citep{chernozhukov2018double}. Here, we describe a data-splitting and cross-fitting technique to help us circumvent this issue and provide valid inference on $\hat{C}_{X \rightarrow Y}$, thereby filling a gap in literature \citep{daniusis2012inferring}. 

Let $\mathcal{D} = \left\{(X_1, Y_1), \ldots, (X_{2n}, Y_{2n}) \right\}$ be a random sample drawn from a bivariate distribution $f_{{XY}}$ with marginal $f_X$ for $X$ and $f_Y$ for $Y$. Since we do not have knowledge of $f_X$ or $f_Y$, we invoke a data-splitting and cross-fitting technique to estimate the underlying density functions as well as the relevant entropy terms. That is, we first split the available data $\mathcal{D}$ into two equal-sized but disjoint sets denoted by $\mathcal{D}_1 := \left\{(X_1, Y_1), \ldots, (X_{n}, Y_{n}) \right\}$, and $\mathcal{D}_2 := \left\{(X_{n+1}, Y_{n+1}), \ldots, (X_{2n}, Y_{2n}) \right\}.$ Using one data split $\mathcal{D}_1$, we obtain estimates of the marginal density functions $\hat{f}_{X; 1}$ and $\hat{f}_{Y; 1}$ by the SCE method described in Section \ref{subsec:density}. The estimated density functions are evaluated for data belonging to the second data split $\mathcal{D}_2$ to obtain the following estimates of marginal entropies $\widehat{H_2}(X) = -  n^{-1}\sum_{j=1}^{n} \ln \left(\hat{f}_{X; 1} \left( {X}_{n + j} \right) \right)$ and $\widehat{H_2}(Y) = - n^{-1}\sum_{j=1}^{n} \ln \left(\hat{f}_{Y; 1} \left( {Y}_{n + j} \right) \right)$. 
Interchanging the roles of data splits $\mathcal{D}_1$ and $\mathcal{D}_2$, by a similar procedure, we obtain the estimated densities $\hat{f}_{X; 2}$ and $\hat{f}_{Y; 2}$. The estimated density functions are evaluated for data belonging to data split $\mathcal{D}_1$ to obtain the estimated entropies $\widehat{H_1}(X) = - n^{-1}\sum_{j=1}^{n} \ln \left(\hat{f}_{X; 2} \left( {X}_{j} \right) \right)$ and $\widehat{H_1}(Y) = - n^{-1}\sum_{j=1}^{n} \ln \left(\hat{f}_{Y; 2} \left( {Y}_{j} \right) \right)$. 
Taking an average of the two sets of estimates, we obtain the so-called ``cross-fitted'' estimates of the marginal entropies $ \hat{H}(X) = \left\{\widehat{H_1}(X) + \widehat{H_2}(X) \right\}/2$ and $\hat{H}(Y) =  \left\{\widehat{H_1}(Y) + \widehat{H_2}(Y) \right\}/2$. Next, we define oracle estimators of $H_X$, and $H_Y$ for each of the data splits:
\begin{equation}
\footnotesize
\begin{aligned}
H_{0; 1}(X) = - \frac{1}{n} \sum_{j=1}^{n} \ln \left({f}_{X} \left( X_{j} \right) \right),  \text{ and } H_{0; 2}(X) = - \frac{1}{n}\sum_{j=1}^{n} \ln \left({f}_{X} \left( X_{n+j} \right) \right),
\end{aligned}
\end{equation} with respect to $\mathcal{D}_1$ and 
\begin{equation}
\footnotesize
    \begin{aligned}
 H_{0; 1}(Y) = - \frac{1}{n}\sum_{j=1}^{n}  \ln \left({f}_{Y} \left( Y_{j} \right) \right),  \text{ and }   H_{0; 2}(Y) = - \frac{1}{n} \sum_{j=1}^{n}  \ln \left({f}_{Y} \left( Y_{n+j} \right) \right), 
\end{aligned}
\end{equation} with respect to $\mathcal{D}_2$. The quantities so obtained are averaged to obtain the ``cross-fitted oracle estimates'' given by $H_{0}(X) = \left\{H_{0; 1}(X) + H_{0; 2}(X)\right\}/{2}$, $H_{0}(Y) =  \left\{H_{0; 1}(Y) + H_{0; 2}(Y)\right\}/{2}$.

The following two theorems establish both consistency and asymptotic normality of the cross-fitted estimates. In addition to Assumptions \ref{ass:th_1_1} -- \ref{ass:th_1_3}, we impose the following assumption that is needed for Theorems \ref{thm:app_consistency} and \ref{them:op-1} to hold. 
\begin{assumption}
\label{ass:th_1_4}
    Let the densities $f_X$ and $f_Y$ be bounded away from zero and infinity on their support. 
\end{assumption} Similar assumptions are made by \cite{Moon1995, Audibert2007}.
\begin{theorem}\label{thm:app_consistency}
Let Assumptions \ref{ass:th_1_1} -- \ref{ass:th_1_4} hold when estimating $\hat{f}_X$ or $\hat{f}_Y$ using data splits $\mathcal{D}_1$ and $\mathcal{D}_2$.  
We implement the data-splitting and cross-fitting procedure described in Section \ref{subsec:split_fit}. Then, the cross-fitted estimate is strongly consistent, i.e., $\hat{C}_{X \rightarrow Y} \overset{a.s.}{\rightarrow} {C}_{X \rightarrow Y}$ as $n \rightarrow \infty$. \end{theorem}

\begin{theorem}\label{them:op-1}
Let Assumptions \ref{ass:th_1_1} -- \ref{ass:th_1_4} hold when estimating $\hat{f}_X$ or $\hat{f}_Y$ using data splits $\mathcal{D}_1$ and $\mathcal{D}_2$.  
We implement the data-splitting procedure described in Section \ref{subsec:split_fit} to obtain cross-fitted estimates $\hat{H}(X)$ and $\hat{H}(Y)$ using $\mathcal{D}_1$ and $\mathcal{D}_2$. Then, we have the following:
\begin{equation} \label{eq:thm2_eq1A}
    \sqrt{n}
\begin{pmatrix}
\hat{H}(X) - H_{0}(X) \\
\hat{H}(Y) -  H_{0}(Y)
\end{pmatrix}\overset{\mathcal{P}}{\rightarrow}  {0}, \text{ as } n \rightarrow \infty.
\end{equation}
\end{theorem}

We refer the reader to the supplement for proofs of Theorem \ref{thm:app_consistency} and Theorem \ref{them:op-1}.

\begin{lemma}
\label{thm:slutsky2}
By the multivariate central limit theorem, assuming $\mathbb{V} \left[\log \left(f_X(X) \right) \right] < \infty$ and $\mathbb{V} \left[\log \left(f_Y(Y) \right) \right] < \infty$, we have $\left(H_0(X), H_0(Y) \right)^\prime$ jointly converge in distribution, given by
\begin{equation*}
\label{eq:thm2_eq1}
    \sqrt{n}
\begin{pmatrix}
H_{0}(X) - H(X) \\
H_{0}(Y) - H(Y)
\end{pmatrix}
\overset{\mathcal{D}}{\rightarrow}  N\left( \mathbf{0}, \Sigma \right), \text{ as } n \rightarrow \infty, 
\end{equation*} where $\Sigma$ is the $2 \times 2$ dispersion matrix of $\left(H_0(X), H_0(Y) \right)^\prime$.
\end{lemma} 
Using Theorem \ref{them:op-1} and Lemma \ref{thm:slutsky2}, we get the following corollary.
\begin{corollary}
\label{thm:asymp_c}
Let Assumptions \ref{ass:th_1_1} -- \ref{ass:th_1_4} hold when obtaining $\hat{C}_{X \rightarrow Y}$ using cross-fitted estimates from data splits $\mathcal{D}_1$ and $\mathcal{D}_2$.  We make use of Lemma \ref{thm:slutsky2} to note that $\sqrt{n}\left(\hat{C}_{X \rightarrow Y} - {C}_{X \rightarrow Y} \right) \overset{\mathcal{D}}{\rightarrow} N(0, \sigma_C^2), \text{ as } n \rightarrow \infty,$ 
where 
$\sigma_C^2$ denotes the asymptotic variance of the estimate $\hat{C}_{X \rightarrow Y}$ and can be estimated by the Monte Carlo technique, given density estimates $\hat{f}_{X}$ and $\hat{f}_Y$.
\end{corollary}
We refer the reader to the supplement for the proof of Corollary \ref{thm:asymp_c}. {See the supplement on how to estimate $\hat{\sigma}^2_C$ using data-splitting and cross-fitting \citep{chernozhukov2018double}.}

\subsection{Testing for asymmetry in GEMs using $C_{X \rightarrow Y}$}
\label{subsec:asymp_ci}
The one-sided hypothesis in \ref{eq:pointwise} postulates a putative directionality from $X$ to $Y$ for $g \in \mathcal{G}_{-}$. This null is ``protected" with a high (\textit{say}, 95\%) confidence and will be rejected if $\hat{C}_{X \rightarrow Y}$ is significantly bigger than zero.  The asymptotic normality established in Corollary~\ref{thm:asymp_c} is the theoretical basis for the proposed hypothesis testing method based on $\hat{C}_{X \rightarrow Y} = \hat{H}(X) - \hat{H}(Y)$. We construct a one-sided $95\%$ asymptotic confidence interval (CI) of $\hat{C}_{X \rightarrow Y}$, denoted by $(\hat{L}, \infty)$ where $\hat{L}$ is the estimated lower bound. Consequently, if $\hat{L} \leq 0$, we would reject the null hypothesis and conclude that the postulated directionality is disproved by the data at hand.} For $g \in \mathcal{G}_{+}$, we would test the one-sided hypothesis in \ref{eq:pointwise2}, given by $H_0: C_{X \rightarrow Y} < 0$.

\section{Simulation studies}
\label{sec:sims}
We now report the findings of extensive simulation studies that provide empirical evidence supporting the validity of $\hat{C}_{X \rightarrow Y}$ to infer GEM-induced asymmetry in bivariate $(X, Y)$.  

\subsection{Behaviour of {${C}_{X \rightarrow Y}$} under {GEM}}
\label{subsec:sim_strong}
In this simulation, we assess the capacity of ${C}_{X \rightarrow Y}$ for measuring the induced asymmetry from a hypothesized GEM in the strong asymmetry framework when the postulate of Assumption \ref{ass:Iden1}  is satisfied. We generate Monte Carlo simulated datasets from GEM models with or without noise disturbance, each consisting of $1000$ $i.i.d.$ samples of pairs, where \textit{GF} $g$ satisfies Assumption \ref{ass:Iden1} under $X \sim U(0, 1)$. We set $g(x)$ to be one of the following functions: $x^{1/3}, x^{1/2}, x^2, x^3, \exp(x),$ and $\sin(\pi x/2)$. Note that the first four GFs have contracting dynamics on $[0, 1]$, while the last two have expanding dynamics. For each setting, the experiments were repeated $R = 250$ times to gauge Monte Carlo approximation errors from a set of empirical values $\{\tilde{C}^{r}_{X \rightarrow Y} \}_{r = 1}^{R}$.  We report the mean along with bottom and upper $2.5^\text{th}$ percentiles of the empirical $\tilde{C}_{X \rightarrow Y}$ in Table \ref{tab:strong}. 
\begin{table}[htp]
\centering
\caption{Examining mean $\tilde{C}_{X \rightarrow Y} \text{ and } (2.5\%,\ 97.5\%)$ {percentiles} in the GEM framework. We consider uniformly distributed $X$ and six choices of $g$. The first row represents the behaviour of ${C}_{X \rightarrow Y}$ in the noise-free GEM given by $Y = g(X)$ whereas the subsequent rows examine shifts in the behaviour of ${C}_{X \rightarrow Y}$ in the NPGEM given by $Y^* = g(X) + \epsilon$, where $\mathbb{V}(\epsilon) = \sigma$.}
\label{tab:strong}
\singlespacing
\footnotesize
\setlength\tabcolsep{0pt}
\begin{tabular*}{\textwidth}{@{}c@{\extracolsep{\fill}}c@{\extracolsep{\fill}}c@{\extracolsep{\fill}}cccc@{}}
\toprule
 \multirow{2}{*}{$\sigma$}  & \multicolumn{4}{c}{$g(x)$ with contracting dynamics} & \multicolumn{2}{c}{$g(x)$ with expanding dynamics} \\ \cmidrule{2-5}  \cmidrule{6-7} \\
  & ${x^{1/3}}$ & ${x^{1/2}}$ & ${x^2}$ & ${x^3}$ & ${\exp(x)}$ & ${\sin(\pi x/2)}$\\
\midrule
 \multirow{2}{*}{$0$} & 0.51 & 0.27 & 0.26 & 0.65 & -0.45 & -0.57\\
  & (0.46, 0.55) & (0.24, 0.30) & (0.19, 0.33) & (0.54, 0.78) & (-0.50, -0.41) & (-0.64, -0.50) \\ 
  \midrule 
   \multirow{2}{*}{$0.10$} & 0.28 & 0.12 & -0.02 & 0.08  & -0.50 & -0.82\\
  & (0.24, 0.32) &  (0.08, 0.15) & (-0.06, 0.02) & (0.04, 0.13) & (-0.54, -0.47) & (-0.92, -0.67) \\ 
  \midrule 
 \multirow{2}{*}{$0.20$} &   -0.00  & -0.10  & -0.23 & -0.18  & -0.61  & -0.86 \\
  & (-0.05, 0.04)& (-0.15, -0.06) &(-0.27, -0.20)& (-0.23, -0.14) & (-0.64, -0.58) &(-0.91, -0.81)\\ 
\bottomrule
\end{tabular*}
\end{table}  Note that here the empirical distribution of $\tilde{C}_{X \rightarrow Y}$ may be used to study behaviour of the asymmetry coefficient under each given GEM.  Further, density estimates of $\tilde{C}_{X \rightarrow Y}$ in Supplementary Figure 1 show the behaviour of the asymmetry coefficient under each NPGEM described in Section \ref{sec:strong_cont}. In all six cases, the average $\tilde{C}_{X \rightarrow Y}$ reflects  the induced asymmetry under each noise-free GEM model. These numerical results confirm the theoretical properties and insights in Section \ref{sec:strong_cont}.

\subsection{Coverage probability, bias and standard error of $\hat{C}_{X \rightarrow Y}$}
\label{subsec:sim1}

We use a simulation experiment to show that our cross-fitting estimation and inference technique can help estimate the proposed  asymmetry coefficient with small error and appropriate coverage probability. We simulate data from $X \sim f_X$ with a present entropy $H(X)$ and generate $Y = g(X)$ in a GEM via a bijective function $g$, with density $f_Y$ and entropy $H(Y)$. We intentionally set $f_X$ and $g$ in a way that the induced population-level parameter ${C}_{X \rightarrow Y} = H(X) - H(Y)$ is positive.  We consider two cases that have both closed-form distributions so that the exact true \textcolor{black}{asymmetry coefficient} is known. (i) $X \sim \text{Lognormal}(5, 1)$ and $g(t) := \log(t)$, which implies $Y \sim \text{N}(5, 1)$ and the true coefficient $=5$; and (ii) $X \sim \text{Exp}(\mu = 1)$ and $g(t) := t^{2/3}$, which implies $Y \sim \text{Weibull}(\text{scale = }1, \text{shape = }3/2)$ and the true coefficient $=0.213$. 
\begin{table}[ht]
\caption{Examining $\hat{C}_{X \rightarrow Y}$: absolute bias (A.Bias), empirical standard error (ESE), the asymptotic SE (ASE) and coverage probability (CP) for different sample sizes $n \in \{250, 500, 750 \}$ under two cases: (i) $X \sim \text{Lognormal}(5, 1)$ and $Y \sim \text{N}(5, 1)$ and (ii) $X \sim \text{Exp}(\text{mean} = 1)$ and $Y \sim \text{Weibull}(\text{scale = }1, \text{shape = }3/2)$.}
\label{tab:coverage}
\singlespacing
\begin{tabular*}{\textwidth}{l@{\extracolsep{\fill}}cccccc}
\toprule
& \multicolumn{3}{c}{Case (I)} & \multicolumn{3}{c}{Case (II)} \\
\cmidrule{2-4} \cmidrule{5-7}
& $n = 250$  & $n = 500$ & $n = 750$ & $n = 250$  & $n = 500$  & $n = 750$ \\
\midrule
A.Bias               & 0.095    & 0.058   & 0.048   & 0.082    & 0.062    & 0.049   \\
ESE  & 0.112    & 0.073   & 0.060   & 0.087    & 0.064    & 0.051   \\
ASE & 0.114    & 0.084   & 0.070   & 0.102    & 0.074    & 0.061   \\ 
CP   & 0.945    & 0.980   & 0.965   & 0.935    & 0.935    & 0.960   \\ \bottomrule
\end{tabular*}
\end{table} 
Our simulation is set up as follows: we vary the sample size $n \in \left\{250, 500, 750 \right\}$; in each case, we obtain both estimate $\hat{C}_{X \rightarrow Y}$ and asymptotic variance $\hat{\sigma}_{C}^2$ according to the formula given in Corollary \ref{thm:asymp_c}. Then, we construct a $95\%$ asymptotic confidence interval, say $\left(\hat{L}_C, \hat{U}_C \right)$. Repeating this procedure over $R = 200$ times, we obtain a set of estimates $\left\{\hat{C}_{X \rightarrow Y}(r), \hat{\sigma}^2_C(r) \right\}_{r = 1}^{R}$ as well as the corresponding $95\%$ asymptotic confidence intervals, $\left\{ \left(\hat{L}_C(r), \hat{U}_C(r) \right)\right\}_{r=1}^R$. 
Table \ref{tab:coverage} lists the empirical mean, absolute bias, and standard error of the asymmetry coefficient estimates as well as the average of the estimated asymptotic variances. In addition, we report the empirical coverage probability, defined as the proportion of $95\%$ asymptotic confidence intervals covering the true  {asymmetry coefficient} parameter. In all the cases considered, our proposed methodology yields an estimate that has low estimation error. The empirical standard error and average asymptotic standard error are close to each other, confirming the validity of the asymptotic normality given in Corollary~\ref{thm:asymp_c}. More importantly, the coverage probability is close to the nominal $95\%$ level in both scenarios, so the proposed cross-fitting inference is numerically stable and trustful.  In summary, these two simulation experiments clearly confirms the large-sample theoretical results given Theorem~\ref{thm:app_consistency} and Corollary~\ref{thm:asymp_c}.

\begin{table}[ht]
\caption{Examining $\hat{C}_{X \succ Y}$: absolute bias (A.Bias), empirical standard error (ESE), the asymptotic SE (ASE) and coverage probability (CP) for different sample sizes $n \in \{250, 500, 750 \}$ under two cases: (i) $X \sim \text{Lognormal}(5, 1)$ and $Y \sim \text{N}(5, 1)$ and (ii) $X \sim \text{Exp}(\text{mean} = 1)$ and $Y \sim \text{Weibull}(\text{scale = }1, \text{shape = }3/2)$.}
\label{tab:coverage}
\begin{tabular*}{\textwidth}{l@{\extracolsep{\fill}}cccccc}
\toprule
& \multicolumn{3}{c}{Case (I)} & \multicolumn{3}{c}{Case (II)} \\
\cmidrule{2-4} \cmidrule{5-7}
& $n = 250$  & $n = 500$ & $n = 750$ & $n = 250$  & $n = 500$  & $n = 750$ \\
\midrule
A.Bias               & 0.095    & 0.058   & 0.048   & 0.082    & 0.062    & 0.049   \\
ESE  & 0.112    & 0.073   & 0.060   & 0.087    & 0.064    & 0.051   \\
ASE & 0.114    & 0.084   & 0.070   & 0.102    & 0.074    & 0.061   \\ 
CP   & 0.945    & 0.980   & 0.965   & 0.935    & 0.935    & 0.960   \\ \bottomrule
\end{tabular*}
\end{table} 

\begin{remark}
{We adopt a 50/50 split for our data, a common and robust choice recommended in the cross-fitting literature \citep{chernozhukov2018double}. This approach provides a balance, ensuring both data folds are sufficiently large for stable nuisance density estimation and subsequent parameter inference. To justify the use of a 50/50 data-splitting ratio, we conducted a simulation study for Cases (I) and (II) under varying sample sizes  to assess the performance of the estimator under various unbalanced splits. The results, presented in supplementary Table 2, show that the $50/50$ split is optimal {in the sense that it} yields the lowest absolute bias and a coverage probability that is closest to the nominal 95\% level. As the split becomes more unbalanced (e.g., 70/30, 90/10), the bias steadily increases and the coverage probability deteriorates significantly. This provides strong empirical support for using the 50/50 split in our main analysis.}
\end{remark}

\subsection{Comparison of performance of $\hat{C}_{X \rightarrow Y}$ with other causal discovery methods}

We compare the accuracy power of our method against three competing methods. The competing methods are the Additive Noise Model (ANM) \citep{mooij_2016}, the Conditional Distribution Similarity (CDS) \citep{Fonollosa2019}, and the Regression Error-based Causal Inference (RECI) \citep{Blbaum2019} approaches. We use the Causal Discovery Toolbox \citep{Kalainathan} to implement these methods, which are used with default parameters supplied by the toolbox.

\subsubsection{Simulated data}

We consider uniformly distributed $X$ and six choices of bijective $g$. We generate $Y = g(X) + \epsilon$, where $\mathbb{V}(\epsilon) = \sigma$ and $\text{Cov}(X, \epsilon) = \rho$. We draw samples of size $n = 1000$ repeatedly for $r = 250$ times and report the proportion of times each method is able to detect the correct directionality $X \rightarrow Y$ and present our findings in Table \ref{tab:other_sim}, which reveals that our method consistently outperform all other methods across all but one simulated data setting. 

\subsubsection{Benchmark data}

We compared the accuracy of the four competing methods using the publicly available \texttt{CauseEffectPairs} benchmark that consists of data for $k = 99$ different cause-effect pairs selected from $37$ data sets from various domains \citep{mooij_2016}. Of the $99$ pairs, ANM was able to correctly detect $51$ pairs and returned inconclusive findings for $18$ pairs, while $\hat{C}_{X \rightarrow Y}$ was able to detect $58$ pairs (under contracting dynamics) and 35 pairs (under expanding dynamics), and was inconclusive for $7$ pairs. The CDS and RECI approaches proved more successful, correctly detecting $67$ and $62$ pairs respectively with no inconclusive findings.

\begin{table}[htp]
\centering
\caption{Comparing performance of $\hat{C}_{X \rightarrow Y}$ with competing pairwise causal discovery methods. We consider uniformly distributed $X$ and six choices of bijective $g$. We generate $Y = g(X) + \epsilon$, where $\mathbb{V}(\epsilon) = \sigma$ and $\text{Cov}(X, \epsilon) = \rho$. We draw samples of size $n = 1000$ repeatedly for $r = 250$ times and report the proportion of times each method is able to detect the correct directionality $X \rightarrow Y$.}
\label{tab:other_sim}
\setlength\tabcolsep{0pt}
\begin{tabular*}{\textwidth}{@{}c@{\extracolsep{\fill}}c@{\extracolsep{\fill}}c@{\extracolsep{\fill}}cccc@{}}
\toprule
\multirow{2}{*}{$g(x)$} & \multirow{2}{*}{$\sigma$} &  \multirow{2}{*}{$\rho$} & \multicolumn{4}{c}{Accuracy} \\  \cmidrule{4-7}
                        &                           &                          &  $\hat{C}_{X \rightarrow Y}$ & ANM & CDS &REI   \\ \midrule
\multirow{3}{*}{$x^{1/3}$} & 0        & 0      & $\mathbf{1.00}$&0.22&0.00&0.00                     \\
                   & 0.05     & 0.10    & $\mathbf{1.00}$&0.24&0.00&0.00                     \\
                   & 0.05     & 0.60    & $\mathbf{1.00}$&0.21&0.00&0.00                     \\ \midrule
\multirow{3}{*}{$x^{1/2}$} & 0        & 0      & $\mathbf{1.00}$&0.18&0.00&0.00                     \\
                   & 0.05     & 0.10    & $\mathbf{1.00}$&0.22&0.00&0.00                     \\
                   & 0.05     & 0.60    & $\mathbf{1.00}$&0.18&0.00&0.00                     \\ \midrule
\multirow{3}{*}{$x^{2}$} & 0        & 0      & $\mathbf{1.00}$&0.24&$\mathbf{1.00}$&$\mathbf{1.00}$                     \\
                   & 0.05     & 0.10    & $\mathbf{1.00}$&0.18&$\mathbf{1.00}$&$\mathbf{1.00}$                     \\
                   & 0.05     & 0.60    & $\mathbf{1.00}$&0.20&$\mathbf{1.00}$&$\mathbf{1.00}$                     \\ \midrule
\multirow{3}{*}{$x^{3}$} & 0        & 0      & $\mathbf{1.00}$&0.20&0.00&$\mathbf{1.00}$                     \\
                   & 0.05     & 0.10    & $\mathbf{1.00}$&0.20&0.02&$\mathbf{1.00}$                     \\
                   & 0.05     & 0.60    & $\mathbf{1.00}$&0.18&0.00&$\mathbf{1.00}$                     \\ \midrule
\multirow{3}{*}{$\exp(x)$} & 0        & 0      & $\mathbf{1.00}$&0.22&0.00&$\mathbf{1.00}$                     \\
                   & 0.05     & 0.10    & $\mathbf{1.00}$&0.20&0.44&$\mathbf{1.00}$                     \\
                   & 0.05     & 0.60    & 0.98&0.21&0.05&$\mathbf{1.00}$                     \\ \midrule
\multirow{3}{*}{$\sin(\pi x/2)$} & 0        & 0      & $\mathbf{1.00}$&0.24&$\mathbf{1.00}$&0.00                     \\
                   & 0.05     & 0.10    & $\mathbf{1.00}$&0.26&$\mathbf{1.00}$&0.00                     \\
                   & 0.05     & 0.60    & $\mathbf{1.00}$&0.26&$\mathbf{1.00}$&0.00                     \\ \bottomrule
\end{tabular*}
\end{table}
    
\section{Application} 
\label{sec:rda}

\subsection{Introduction}
We apply the above asymmetry analytic to study the epigenetic relation between DNA methylation (\textit{DNAm}) and blood pressure (\textit{BP}) that is one of the most important risk factors for cardiovascular disease (\textit{CVD}). We use data from $n = 522$ children (including $247$ boys and $275$ girls) aged 10-18 years in the Early Life Exposures in Mexico to Environmental Toxicants (\textit{ELEMENT}) study \citep{Hernandez_Avila_1996}. 
The primary task of scientific interest is to understand potential \textit{CVD} causal pathways, part of which involves examining asymmetry between \textit{DNAm} alterations and \textit{CVD} risks. In this analysis, we examine six candidate genes found to be significantly associated with systolic blood pressure (\textit{SBP}) and diastolic blood pressure (\textit{DBP}) in at least 20 independent studies according to the NHGRI-EBI GWAS Catalog. These six target genes include \textbf{FGF5, HSD11B2, KCNK3, ATP2B1, ARHGAP42} and \textbf{PRDM8}. To investigate their epigenetic roles in \textit{CVD}, we investigate whether $\beta$ values of \textit{DNAm} (specifically, cytosine-phosphate-guanine (CpG) methylation) influence change in \textit{BP} or if the converse is true \citep{dicorpo_2018, hong2023association}.

\subsection{Methods}

{To perform a gene-level analysis, we average {beta-}values of \textit{DNAm} over CpG sites within each gene, which is then normalized across subjects by an affine transformation described in Section \ref{subsec:gcm_setup} to minimize any undue influence of location or scale changes in individual \textit{DNAm} measurements. The same normalization procedure is applied to \textit{SBP} and \textit{DBP} measurements. Following normalization, we apply a rigorous three-step analysis to each gene-sex-\textit{BP} triplet to infer the directional relationship between \textit{DNAm} and \textit{BP} (either \textit{DBP} or \textit{SBP}). First, we determine the plausible causal direction by verifying the functional orthogonality assumption (Assumption 1) for either potential GEMs: \textit{DNAm} $= g(\textit{BP})$ and BP $= h(\textit{DNAm})$ using bootstrap-based orthogonality deviation score described in Algorithm 1 for this purpose. The analysis proceeds only if the bootstrap confidence interval for the deviation score contains zero for exactly one of the two directions. If neither or both directions satisfy the assumption, the result is deemed inconclusive. Second, for the single direction that satisfies functional orthogonality, we characterize the dynamics of its generative function using a bootstrap procedure (see Remark 7  for more details). In brief, we estimate the average log-gradient of the function to classify it as either `contracting' or `expanding' dynamics. This step is crucial as it determines the specific {null} hypothesis to be tested; a contracting function implies a positive asymmetry coefficient ($C_{X \rightarrow Y} > 0$), whereas an expanding function implies a negative one ($C_{X \rightarrow Y} < 0$). Finally, having established a valid direction and the nature of the generative function, we proceed to the formal inference step of our hypothesized GEM. We employ the data-splitting and cross-fitting procedure to compute the asymmetry coefficient ($\hat{C}_{X \rightarrow Y}$) and its corresponding one-sided $95\%$ confidence interval. This allows us to test the directionality hypothesis formulated in the previous step and draw a conclusion about the epigenetic pathway while performing diagnostic checks to verify if the underlying assumptions are satisfied.}

\subsection{Results}

{We applied our three-step analytical pipeline to investigate the directional relationship between \textit{DNAm} and blood pressure (systolic - \textit{SPB} and diastolic - \textit{DBP}) for six candidate genes, stratifying the analysis by sex. Our findings reveal distinct directional pathways for several genes.}

{For genes \textbf{FGF5} and \textbf{HSD11B2}, the analysis yielded strong, consistent evidence for a directional influence from blood pressure to \textit{DNAm}. In all tested scenarios---for both males and females, and for both \textit{SBP} and \textit{DBP}---the functional orthogonality assumption was uniquely satisfied for the BP $\rightarrow$ \textit{DNAm} direction. The generative functions were consistently identified as having contracting dynamics, and the resulting asymmetry coefficient was statistically significant, with the one-sided 95\% confidence interval for $\hat{C}_{\textit{BP} \rightarrow \textit{DNAm}}$ being strictly positive.}

{The results for \textbf{KCNK3} and \textbf{ATP2B1} were more varied. For \textbf{KCNK3}, we identified a significant pathway from \textit{SBP} to \textit{DNAm} in males and from \textit{DBP} to \textit{DNAm} in females. However, the relationship between \textit{SBP} and \textit{DNAm} in females was inconclusive, as the orthogonality assumption was violated. For \textbf{ATP2B1}, significant pathways from both \textit{SBP} and \textit{DBP} to \textit{DNAm} were found in males. In females, the \textit{SBP}-\textit{DNAm} relationship was inconclusive due to a similar assumption violation, and no significant directionality was found with \textit{DBP}.}

{Next, for genes \textbf{ARHGAP42} and \textbf{PRDM8}, our analysis did not detect any significant directional asymmetry. Across all sex and blood pressure combinations for these two genes, the estimated asymmetry coefficient was not statistically distinguishable from zero, providing no evidence to support a directional relationship in either direction.}

{Finally, given that noise contamination can particularly affect inference for GFs with contracting dynamics, we also performed a diagnostic check to ensure that the estimated signal-to-noise ratio for each significant pathway was well within the theoretical tolerance bound established in Section \ref{sec:strong_cont}. Our investigations reveal that the noise perturbation has little influence on the positive results drawn from our inference method.}

\subsection{Discussion}
Our inference method unveils novel pathways from \textit{BP} to \textit{DNAm} 
for two genes \textbf{FGF5} and \textbf{HSD11B2}. \textbf{FGF5} is reported in \cite{FGF5} as a key gene in several animal studies that increases myocardial blood flow and function, decrease myocyte apoptosis, and increase myocyte number after gene transfer of the growth factor. However, there are no available results of its effects in human cardiovascular disease. Our analysis is the first to unveil a pathway from changes in \textit{BP} to regulate \textit{DNAm} of this key \textit{CVD} gene in a human cohort study. Gene \textbf{HSD11B2} was reported in \cite{HSD11B2} to be associated with obesity-related cardiovascular risk factors, e.g. type II diabetes and hypertension. Hypertension is often associated with chronic inflammation and oxidative stress, both of which can trigger an imbalance in the immune system.
This can, in turn, modulate a person's methylation patterns.  In summary, these new findings confer an added sense of directionality in the study of \textit{BP} variation and epigenetic biomarkers, paving the way for future advancements in individualized risk assessments and even therapeutic targets. 

\section{Concluding remarks}
Asymmetry is an inherent property of bivariate associations and therefore must not be ignored~\citep{zheng2012generalized}. 
In this paper, we present a new methodology in the framework of generative exposure mapping (GEM) models so that the induced asymmetry between two random variables $X$ and $Y$ is captured by an information-theoretic coefficient $C_{X \rightarrow Y}$. 
Utilizing this asymmetry measure, we develop a cross-fitting inference that enables us to examine a certain hypothesized directionality with proper uncertainty quantification.  The intrinsic linkage between information theory and an experiment with a uniformly distributed exposure provides an interesting insight into asymmetry from the perspective of causality, where we argue that the induced asymmetry from a GEM model may be regarded as a low-dimensional imprint of causality.  Further, simulation studies reveal that the asymmetry measure $C_{X \rightarrow Y}$ can detect different dynamics of outcomes under different GEMs without actually estimating the mapping functions, and such capacity prevails in noise-perturbed GEM models. It is worth noting that such an extension requires one assumption of the contamination error being independent of the exposure, although empirical studies show robustness of the asymmetry coefficient to endogeneity in contaminated GEM models. {A potential area of future work would involve addressing endogeneity in contaminated GEM models with correlations between the errors and exposure \citep{breunig2021testability}, in which the instrumental variable technique is worth exploring. This extension would require identifying a valid instrument $W$ that is correlated with $X$ but independent of the error $\epsilon$. In such a setup, we would first use the instrument $W$ to model $X$, analogous to the ``first stage" of a Two-Stage Least Squares regression \citep{angrist1995identification}. Hence, one could isolate the ``clean" component of the exposure (i.e., ${X}_{IV} = f(W)$) and redefine the asymmetry coefficient using this ``clean'' component, for instance, as $C^{IV}_{X \rightarrow Y} = H({X}_{IV}) - H(Y)$. This IV-based approach would, however, require significant new theoretical development to extend our {current} framework, and to derive the statistical properties of this new coefficient.}

{Another important extension is the incorporation of covariates (or modifiers), $\boldsymbol{Z}$, into the GEM framework. This presents several challenges that future work must address. First, it will require theoretical effort to extend the concept of functional orthogonality (as formalized by Assumption 1)  to a conditional setting. The simple independence between the input density and the generative function  must be re-conceptualized given $\boldsymbol{Z}$. Consequently, a new diagnostic tool would be needed to verify this conditional orthogonality. The current ``orthogonality deviation score" (see Algorithm \ref{alg:ortho_check})  is only defined for the bivariate case and would be insufficient. Finally, the asymmetry coefficient itself would need to be refined. A promising approach would be to develop a ``conditional asymmetry coefficient'', likely based on conditional entropies, to properly isolate the ``low-dimensional imprint of causality"  after accounting for the influence of $\boldsymbol{Z}$.}

Several further applications and extensions of the proposed  $C_{X \rightarrow Y}$ measure and cross-fitting inference method are worth exploring, including the extension of GEM-induced asymmetry in functional or longitudinal variables; for example, the found asymmetries between \textit{DNAm} and \textit{BP} in our data analysis would become more reliable if repeated measurements of \textit{BP} are available. 
In conclusion, our framework may be used either as a discovery tool, thereby aiding practitioners in improving the rigor and reproducibility of scientific research.

	\section*{Supplementary Materials}
    \begin{itemize}
        \item Section I (Proofs): consistency results, asymptotic normality, and convergence properties of the estimators under generative exposure mappings. 
        \item Section II (Behavior in NPGEMs): how $\hat{C}_{X \rightarrow Y}$ behaves under NPGEMs.
        \item Section III (Methylation Data Application): Additional cross-inference results.
        \item Section IV (Data-splitting and Cross-fitting): This section details the technical implementation of the inference procedure and the variance estimation method needed for constructing confidence intervals and hypothesis tests.
    \end{itemize}
	
	\section*{Acknowledgements}
	
	This work is supported by NSF DMS-2113564 and NIH R01ES033656 (for Song), and the University of Michigan Rackham Predoctoral Fellowship (for Purkayastha). This research was further supported in part by the University of Pittsburgh Center for Research Computing and Data, RRID:SCR022735, through the resources provided. Specifically, this work used the HTC cluster, which is supported by NIH award number S10OD028483.
    


\bibhang=1.7pc
\bibsep=2pt
\fontsize{9}{14pt plus.8pt minus .6pt}\selectfont
\renewcommand\bibname{\large \bf References}
\expandafter\ifx\csname
natexlab\endcsname\relax\def\natexlab#1{#1}\fi
\expandafter\ifx\csname url\endcsname\relax
  \def\url#1{\texttt{#1}}\fi
\expandafter\ifx\csname urlprefix\endcsname\relax\def\urlprefix{URL}\fi

\bibliographystyle{chicago}      
\bibliography{SS-template/00_bioblio}   

\end{document}